\documentclass[pre,twocolumn,showpacs,superscriptaddress]{revtex4}

\usepackage{amsmath}
\usepackage{graphicx}
\usepackage{mathrsfs}

\usepackage{amsmath}
\usepackage{amssymb}
\usepackage{graphicx}
\usepackage[british]{babel}

\begin{document}

\title{Bulk-mediated diffusion on a planar surface: full solution}

\author{Aleksei V. Chechkin}
\affiliation{Institute for Theoretical Physics NSC KIPT,
Akademicheskaya st.1, 61108 Kharkov, Ukraine}
\affiliation{Institute for Physics \& Astronomy, University of Potsdam,
D-14476 Potsdam-Golm, Germany}
\author{Irwin M. Zaid}
\affiliation{Rudolf Peierls Centre for Theoretical Physics, University of
Oxford, 1 Keble Road, Oxford OX1 3NP, United Kingdom}
\author{Michael A. Lomholt}
\affiliation{MEMPHYS - Center for Biomembrane Physics, Department of
Physics, Chemistry, and Pharmacy, University of Southern Denmark, Campusvej 55,
5230 Odense M, Denmark}
\author{Igor M. Sokolov}
\affiliation{Institut f{\"u}r Physik, Humboldt Universit{\"a}t
zu Berlin, Newtonstra{\ss}e 15, 12489 Berlin, FRG}
\author{Ralf Metzler}
\affiliation{Institute for Physics \& Astronomy, University of Potsdam,
D-14476 Potsdam-Golm, Germany}
\affiliation{Department of Physics, Technical University of Tampere,
FI-33101 Tampere, Finland}


\date{\today}

\begin{abstract}
We consider the effective surface motion of a particle that intermittently
unbinds from a planar surface and performs bulk excursions. Based on a
random walk approach we derive the diffusion equations for surface and bulk
diffusion including the surface-bulk coupling. From these exact dynamic
equations we analytically obtain the propagator of the effective surface
motion. This approach allows us to deduce a superdiffusive, Cauchy-type
behavior on the surface, together with exact cutoffs limiting the Cauchy
form. Moreover we study the long-time dynamics for the surface motion.
\end{abstract}

\maketitle

\section{Introduction}

\begin{figure}
\includegraphics[width=8cm]{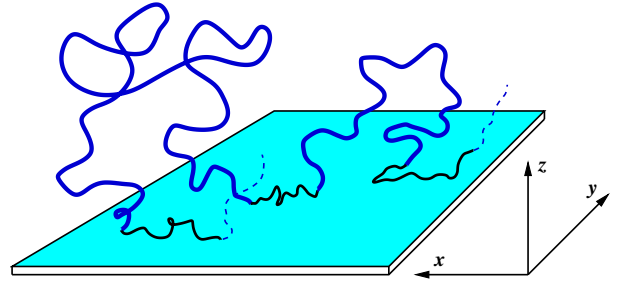}
\caption{Schematic of the bulk mediated surface diffusion: The thinner
(black) lines show the motion along the planar surface with surface
diffusivity $D_s$. Following dissociation from the surface, thicker (blue)
lines depict excursions into
the bulk volume, in which the diffusion constant is $D_b$. Eventually, the
particle rebinds to the surface.}
\label{scheme}
\end{figure}

Interfaces and the interaction of particles with them play a crucial role on
small scales in biology and technology. For instance, biopolymers such as
proteins or enzymes diffusing in biological cells intermittently bind to
cellular membranes, or individual bacteria forming a biofilm on a surface
use bulk excursions to efficiently relocate. Similarly the exchange between
a liquid phase with a solid surface is an important phenomenon in the
self-assembly of surface layer films and is a ubiquitous process in emulsions.
This bulk-mediated surface diffusion, schematically shown in Fig.~\ref{scheme},
was previously analyzed in terms of scaling arguments and simulations
\cite{bychuk,bychuk1,revelli,fatkullin}, and was unveiled in field cycling NMR
experiments in porous glasses \cite{stapf}. Moreover, effects of bulk-surface
interchange were reported on proton transport across biological membranes
\cite{membrane}. Recent studies are concerned with effects of bulk-surface
exchange on reaction rates in interfacial systems \cite{benichou} and with
surface diffusion of coppper atoms in nanowire fabrication \cite{nanowire}.

The remarkable finding of the bulk mediated surface diffusion model is that
the effective surface motion is characterized by a Cauchy propagator
\cite{bychuk,bychuk1,revelli,fatkullin}
\begin{equation}
n_s(\mathbf{r},t)\simeq\frac{c^{1/2}t}{2\pi(r^2+ct^2)^{3/2}},
\end{equation}
where $\simeq$ denotes a scaling property ignoring multiplicative constants,
and $c$ is a dimensional factor. The associated stochastic transport is of
superdiffusive nature \cite{bychuk,bychuk1,revelli,fatkullin,stapf,report},
\begin{equation}
\langle\mathbf{r}^2(t)\rangle_s\simeq t^{3/2}.
\end{equation}
Here we present a
strictly analytical approach to this process. Our findings corroborate
the previous scaling results for superdiffusion, however, we also derive the
cutoffs to this behavior: at sufficiently long distances, the Cauchy propagator
turns over to a Gaussian wing. Moreover at longer times the effective surface
diffusion becomes subdiffusive, due to the fact that the particle spends less
and less time on the surface. Normalized to the time-dependent surface coverage
the effective surface diffusion turns over from superdiffusion to normal
diffusion.

\section{Derivation from a discrete random walk process}

We start with a derivation of the coupling between surface and bulk in a
discrete random walk process along the $z$ coordinate perpendicular to the
surface (Fig.~\ref{scheme}). Let $N_i$ with $i=1,2,\ldots$ denote the number
of particles at site $i$ of this one-dimensional lattice with spacing $a$.
The number of particles on the surface at lattice site $i=0$ are termed
$\mathcal{N}_0$. The exchange of particles is possible only via nearest
neighbor jumps. Each jump event between bulk sites $i=1,2,\ldots$ is
associated with the typical waiting time $\tau$. For the exchange between the
surface and site $i=1$ we then have the following law
\begin{equation}
\label{exch}
\frac{d\mathcal{N}_0(t)}{dt}=\frac{1}{2\tau}N_1-\frac{1}{\tau_{\mathrm{des}}}
\mathcal{N}_0,
\end{equation}
where $\tau_{\mathrm{des}}$ is the characteristic time for desorption from
the surface. The probability of adsorption to the surface here is one, i.e.,
a particle adds to the surface population automatically when moving from
site 1 to site 0. The exchange site at $i=1$ between the surface at $i=0$
and the next bulk site $i=2$ is governed by the balance relation
\begin{equation}
\frac{dN_1(t)}{dt}=\frac{1}{\tau_{\mathrm{des}}}\mathcal{N}_0-\frac{1}{\tau}
N_1+\frac{1}{2\tau}N_2.
\end{equation}
Finally, the bulk sites $i=2,3,\ldots$ are governed by equations of the form
\begin{equation}
\frac{dN_2(t)}{dt}=\frac{1}{2\tau}N_1+\frac{1}{2\tau}N_3-\frac{1}{\tau}N_2,
\label{exch1}
\end{equation}
etc. Let us define the number of ``bulk'' particles at the surface site $i=0$
through
\begin{equation}
\label{exch4}
N_0\equiv\frac{2\tau}{\tau_{\mathrm{des}}}\mathcal{N}_0.
\end{equation}
This trick allows us to formulate the exchange equation also for site
$i=1$ in a homogeneous form. Namely, from Eq.~(\ref{exch}) we have
\begin{equation}
\label{exch2}
\frac{d\mathcal{N}_0(t)}{dt}=\frac{1}{2\tau}\left(N_1-N_0\right).
\end{equation}
Moreover, from Eqs.~(\ref{exch1}) we find
\begin{equation}
\label{exch3}
\frac{dN_i(t)}{dt}=\frac{1}{2\tau}\left(N_{i-1}-2N_i+N_{i+1}\right),
\end{equation}
for $i\ge1$.

Let us now take the continuum limit. For that purpose we make a transition
from $\mathcal{N}_0\to n_s$ as the number of surface particles, and $N_i
\to an_b$ for the bulk concentration of particles. The factor $a$ may be
viewed as the lattice constant measuring the distance between successive
sites along the $z$ axis. As the bulk density is a function of $z$ while
the surface particles are all assembled at $z=0$, we need this length
scale $a$ to match dimensionalities between $n_s$ and $n_b$ which are
$[n_s]=\mathrm{cm}^{-2}$ and $[n_b]=\mathrm{cm}^{-3}$. Expansion of
the right hand side of Eq.~(\ref{exch2}) yields the surface-bulk coupling
\begin{equation}
\frac{\partial n_s(t)}{\partial t}=\frac{a}{2\tau}a\left.\frac{\partial n_b
(z,t)}{\partial z}\right|_{z=0}.
\end{equation}
Similarly from Eq.~(\ref{exch3}) we obtain the bulk diffusion equation
\begin{equation}
\frac{\partial n_b(z,t)}{\partial t}=\frac{1}{2\tau}a^2\frac{\partial^2 n_b}{
\partial z^2}.
\end{equation}
Finally, the boundary condition
\begin{equation}
\left.\frac{a}{2\tau}n_b(z,t)\right|_{z=0}=\frac{1}{\tau_{\mathrm{des}}}n_s
\end{equation}
stems from our definition (\ref{exch4}). This completes the description of
the particle exchange between surface and bulk, as well as the diffusion
of bulk particles along the $z$ axis.

We now turn back to the full three-dimensional problem and add to above
equations the $(x,y)$ directions (see Fig.~\ref{scheme}). As the motion
in the perpendicular $z$ direction is fully independent, we may simply
adjust the Laplacian to three dimensions, ending up with the
diffusion equation
\begin{equation}
\label{bulk_diff}
\frac{\partial n_b(x,y,z,t)}{\partial t}=D_b\left(\frac{\partial^2}{\partial
x^2}+\frac{\partial^2}{\partial y^2}+\frac{\partial^2}{\partial z^2}\right).
\end{equation}
with diffusivity $D_b$. For the surface diffusion along the $x$ and $y$
coordinates with $z=0$, we end up with the two-dimensional diffusion
equation with diffusivity $D_s$, plus the bulk-surface exchange term:
\begin{equation}
\label{surf_diff}
\frac{\partial n_s(x,y,t)}{\partial t}=D_s\left(\frac{\partial^2}{\partial
x^2}+\frac{\partial^2}{\partial y^2}\right)n_s+\left.D_b\frac{\partial n_b}{
\partial z}\right|_{z=0}.
\end{equation}
These two equations are valid in the range $0\le z<\infty$ and $-\infty<x,y<
\infty$, and are supplemented by the initial condition
\begin{equation}
\label{surf_init}
n_s(x,y,t)\Big|_{t=0}=n_0\delta(x)\delta(y)
\end{equation}
indicating that initially the particles are all concentrated on the surface
at $x=y=0$. Moreover we observe the boundary condition for $n_b$,
\begin{equation}
\label{bound}
n_b(x,y,z,t)\Big|_{z=0}=\mu n_s(x,y,t),
\end{equation}
and the bulk initial condition
\begin{equation}
n_b(x,y,z,t)\Big|_{t=0}=0.
\end{equation}
In what follows, for simplicity of notation we use a unit initial concentration
$n_0$.

The surface and bulk diffusivities can be expressed in terms of the lattice
constant $a$ and the typical waiting times between jumps in the bulk, $\tau$,
and on the surface, $\tau_s$, through
\begin{equation}
D_b\equiv\frac{a^2}{6\tau},
\end{equation}
and
\begin{equation}
D_s\equiv\frac{a^2}{4\tau_s},
\end{equation}
respectively. In the continuum limit, both $a$ and $\tau$ (or $\tau_s$) tend
to zero such that the diffusion constants remain finite. In many physical
systems bulk diffusion is considerably faster, such that $D_s\ll D_b$. Above
we also introduced the coupling parameter
\begin{equation}
\mu\equiv\frac{a}{D_b\tau_{\mathrm{des}}}
\end{equation}
of physical dimension $[\mu]=1/\mathrm{cm}$. Small values of $\mu$ at fixed
$a$ and $D_b$ correspond to slow bulk-surface exchange.

For consistency we derive the overall number of particles. To this end we
define the number particles on the surface, $N_s(t)$, and in the bulk,
$N_b(t)$ through the relations
\begin{equation}
N_s(t)=\int_{-\infty}^{\infty}dx\int_{-\infty}^{\infty}dy\,n_s(x,y,t)
\end{equation}
and
\begin{equation}
N_b(t)=\int_0^{\infty}dz\int_{-\infty}^{\infty}dx\int_{-\infty}^{\infty}dy\,
n_s(x,y,t).
\end{equation}
From integration of Eq.~(\ref{surf_diff}) we find
\begin{equation}
\label{surf_numb}
\frac{dN_s(t)}{dt}=D_b\left.\frac{\partial}{\partial z}\int_{-\infty}^{\infty}dx
\int_{-\infty}^{\infty}dy\,n_b(x,y,z,t)\right|_{z=0}.
\end{equation}
Similarly, Eq.~(\ref{bulk_diff}) yields
\begin{eqnarray}
\nonumber
\frac{dN_b(t)}{dt}&=&D_b\int_0^{\infty}dz\frac{\partial^2}{\partial z^2}\int_{
-\infty}^{\infty}dx\int_{-\infty}^{\infty}dy\,n_b(x,y,z,t)\\
&=&-D_b\left.\frac{\partial}{\partial z}\int_{-\infty}^{\infty}dx\int_{-\infty}
^{\infty}dy\,n_b(x,y,z,t)\right|_{z=0}.
\label{bulk_numb}
\end{eqnarray}
Combination of Eqs.~(\ref{surf_numb}) and (\ref{bulk_numb}) produces
\begin{equation}
\frac{d}{dt}\Big(N_s(t)+N_b(t)\Big)=0.
\end{equation}
Thus the overall number of particles is conserved, as it should be.

\section{Solution of the coupled diffusion problem}

To solve the set of coupled equations (\ref{surf_diff}) and (\ref{bulk_diff})
for the specified initial and boundary value problem, we start by defining
the two-dimensional Green's function
\begin{widetext}
\begin{equation}
G_s(x,x',y,y',t)\equiv G_s(x-x',y-y',t)=\frac{1}{\sqrt{4\pi D_st}}\exp\left(-
\frac{(x-x')^2+(y-y')^2}{4D_st}\right).
\end{equation}
Then, the solution of Eq.~(\ref{surf_diff}) becomes
\begin{eqnarray}
\nonumber
n_s(x,y,t)&=&\int dx'\int dy'G_s(x,x',y,y',t)n_s(x',y',t=0)\\
&&+\int_0^tdt'\int dx'
\int dy'G_s(x,x',y,y',t-t')\left(D_b\frac{\partial n_b(x',y',z,t')}{\partial z}
\right)_{z=0}.
\end{eqnarray}
With initial condition (\ref{surf_init}) we thus find
\begin{equation}
\label{surf_fl}
n_s(k_x,k_y,s)=G_s(k_x,k_y,s)+G_s(k_x,k_y,s)\mathscr{L}\left\{
\mathscr{F}\left\{D_b\left(\frac{\partial n_b(x,y,z,t)}{\partial z}\right)_{
z=0};x\to k_x;y\to k_y\right\};t\to s\right\},
\end{equation}
where
\begin{equation}
\label{greens}
G_s(k_x,k_y,s)=\frac{1}{s+D_b\left[k_x^2+k_y^2\right]}
\end{equation}
is the Fourier-Laplace transform of the surface Green's function $G_s$.
Here and in the following we denote the Laplace and Fourier transform of a
function by explicit dependence on the image variables, that is,
\begin{equation}
f(s)=\mathscr{L}\{f(t);t\to s\}=\int_0^{\infty}f(t)e^{-st}dt\quad\mbox{and}
\quad g(k)=\mathscr{F}\{g(x);x\to k\}=\int_{-\infty}^{\infty}g(x)e^{ikx}dx.
\end{equation}
The bulk particle density according to Eq.~(\ref{bulk_diff}) is given by
the formal expression \cite{jaeger}
\begin{equation}
\label{bulk_fl}
n_b(x,y,z,t)=\frac{z}{\left(4\pi D_b\right)^{3/2}}\int_0^tdt'\int dx'\int dy'
\frac{\mu n_s(x',y',t')}{(t-t')^{5/2}}\exp\left(-\frac{(x-x')^2+(y-y')^2+z^
2}{4D_b(t-t')}\right).
\end{equation}
From this expression one can indeed show that, despite the factor $z$, the
coupling equation (\ref{bound}) is fulfilled, see Appendix \ref{app1}. Now,
Eq.~(\ref{surf_fl}) requires the derivative of expression (\ref{bulk_fl})
with respect to $z$, evaluated at $z=0$. To find that expression, we first
differentiate
\begin{equation}
\label{expr}
\frac{\partial n_b}{\partial z}=\frac{\mu}{\left(4\pi D_b\right)^{3/2}}\int_0
^tdt'\int dx'\int dy'\frac{n_s(x',y',t')}{(t-t')^{5/2}}\exp\left(-\frac{(x-
x')^2+(y-y')^2+z^2}{4D_b(t-t')}\right)\left[1-\frac{z^2}{2D_b(t-t')}\right],
\end{equation}
and then calculate its Fourier-Laplace transform
\begin{equation}
\mathscr{L}\left\{\mathscr{F}\left\{D_b\left(\frac{\partial n_b}{
\partial z}\right)_{z=0}\right\}\right\}=\frac{\mu D_b}{\sqrt{4\pi D_b}}
n_s(k_x,k_y,s)\mathscr{L}\left\{\frac{\exp\left(-D_b\left[k_x^2+k_y^2\right]
t\right)}{t^{3/2}}\right\}_{z=0},
\end{equation}
due to the convolution nature of expression (\ref{expr}). The Laplace transform
is evaluated by help of the shift theorem, yielding
\begin{equation}
\mathscr{L}\left\{\frac{\exp\left(-D_b[k_x^2+k_y^2]t\right)}{t^{3/2}}\right\}
=\mathscr{L}\left\{\frac{1}{t^{3/2}}\right\}_{s\to s+D_b[k_x^2+k_y^2]}=
-\sqrt{4\pi\left(s+D_b[k_x^2+k_y^2]\right)}.
\end{equation}
We thus finally obtain
\begin{equation}
\mathscr{L}\left\{\mathscr{F}\left\{D_b\left(\frac{\partial n_b}{
\partial z}\right)_{z=0}\right\}\right\}=-\mu D_b^{1/2}\sqrt{s+D_b\left(
k_x^2+k_y^2\right)}n_s(k_x,k_y,s).
\end{equation}
\end{widetext}
Insertion of this equation into expression (\ref{surf_fl}) delivers the
solution for the surface density in Fourier-Laplace space,
\begin{equation}
n_s(k_x,k_y,s)=\frac{G_s(k_x,k_y,s)}{1+\mu D_b^{1/2}G_s(k_x,k_y,s)\sqrt{s+
D_b\left[k_x^2+k_y^2\right]}},
\end{equation}
where the Fourier-Laplace transform of the Green's function $G_s$ was defined
in Eq.~(\ref{greens}). After some transformations we arrive at the exact closed
form expression
\begin{equation}
\label{result_fl}
n_s(k_x,k_y,s)=\frac{1}{s+D_s\left[k_x^2+k_y^2\right]+\chi\sqrt{s+
D_b\left[k_x^2+k_y^2\right]}},
\end{equation}
with the rescaled coupling parameter
\begin{equation}
\chi\equiv\mu D_b^{1/2}=\frac{a}{D_b^{1/2}\tau_{\mathrm{des}}}
\end{equation}
of dimension $[\chi]=1/\mathrm{sec}^{1/2}$. Relation (\ref{result_fl}) is
the main result of this work, and we now consider the consequences to the
surface motion effected by the bulk mediation.

\section{Effective surface behavior}

We first determine the number of particles on the surface,
\begin{equation}
N_s(t)=\int_{-\infty}^{\infty}\int_{-\infty}^{\infty}n_s(x,y,t)dxdy,
\end{equation}
whose Laplace transform is
\begin{equation}
N_s(s)=n(k_x,k_y,s)\Big|_{k_x=k_y=0}=\frac{1}{s+\chi s^{1/2}}.
\end{equation}
Inverse Laplace transformation then yields the exact expression
\begin{equation}
N_s(t)=e^{\chi^2t}\mathrm{erfc}\Big(\chi t^{1/2}\Big),
\end{equation}
where we use the complementary error function
\begin{equation}
\mathrm{erfc}(z)=\frac{2}{\sqrt{\pi}}\int_z^{\infty}e^{-\xi^2}d\xi=1-
\mathrm{erf}(z).
\end{equation}
At short times $t\ll\chi^{-2}$, this leads to the initial decay
\begin{equation}
\label{numb_short}
N_s(t)\sim1-\frac{2\chi}{\sqrt{\pi}}t^{1/2}
\end{equation}
of the number of surface particles, eventually turning into the long
time behavior
\begin{equation}
\label{numb_long}
N_s(t)\sim\frac{1}{\chi\sqrt{\pi t}}.
\end{equation}
The asymptotic $1/\sqrt{\pi t}$ decay stems from the returning dynamics to the
origin $z=0$ of a Brownian motion along the $z$ coordinate, i.e., it is
proportional to the normalization factor of a one-dimensional Brownian motion.
The additional prefactor $\chi$ rescales time with respect to the efficiency of
the surface-bulk exchange.

We now turn to the surface dynamics, as quantified by the effective mean
squared displacement along the surface. In the Laplace domain,
\begin{eqnarray}
\nonumber
\left<\mathbf{r}^2(s)\right>_s&=&-\nabla^2_{k_x,k_y}n(k_x,k_y,s)\Big|_{k_x=k_y
=0}\\
\nonumber
&=&-\left[\frac{\partial^2}{\partial k^2}+\frac{1}{k}\frac{\partial}{\partial
k}\right]n(k_x,k_y,s)\Big|_{k_x=k_y=0}\\
&=&\frac{4D_s}{s(\sqrt{s}+\chi)^2}+\frac{2\chi D_b}{s^{3/2}(s^{1/2}+\chi)^2},
\end{eqnarray}
where $k=|\mathbf{k}|=\sqrt{k_x^2+k_y^2}$.
From this expression we obtain the limiting behaviors at short and long
times. Thus, we observe that the short time limit $t\ll1/\chi^2$ in Laplace
domain corresponds to $s\gg\chi^2$, such that 
\begin{equation}
\left<\mathbf{r}^2(s)\right>_s\sim\frac{4D_s}{s^2}+\frac{2\chi D_b}{s^{5/2}}.
\end{equation}
This translates into the asymptotic time evolution
\begin{equation}
\label{msd_asymp}
\left<\mathbf{r}^2(t)\right>_s\sim4D_s t\left(1+\frac{2}{3\sqrt{\pi}}\frac{
D_b}{D_s}\left[t\chi^2\right]^{1/2}\right).
\end{equation}
As long as the ratio $D_b/D_s$ is sufficiently large, there is a superdiffusive
component $\left<\mathbf{r}^2(t)\right>_s\sim t^{3/2}$ winning over the normal
surface diffusion proportional to $D_s$. This is exactly the famed bulk mediated
superdiffusion originally obtained from scaling arguments by Bychuk and
O'Shaugnessy \cite{bychuk,bychuk1}. Note that this diffusional enhancement is
accompanied by an almost constant number of surface particles, compare
Eq.~(\ref{numb_short}).

Conversely, at long times $t\gg1/\chi^2$ (or $s\ll\chi^2$ in Laplace domain) we
find
\begin{equation}
\left<\mathbf{r}^2(s)\right>_s\sim\frac{4D_s}{\chi^2s}+\frac{2D_b}{\chi s^{3/2}},
\end{equation}
corresponding to the temporal behavior
\begin{equation}
\label{momlong}
\left<\mathbf{r}^2(t)\right>_s\sim\frac{4D_s}{\chi^2}\left(1+\frac{1}{\sqrt{\pi}}
\frac{D_b}{D_s}\left[t\chi^2\right]^{1/2}\right).
\end{equation}
Somewhat surprisingly, at sufficiently long times the bulk contribution to the
effective mean squared displacement dominates over the surface contribution
for arbitrary ratio $D_b/D_s$, giving rise to \emph{subdiffusive\/} behavior.
This subdiffusion occurs due to the ongoing loss of surface particles into the
bulk, see Eq.~(\ref{numb_long}).

Instead of considering the surface mean squared displacement $\left<\mathbf{r
}^2(t)\right>_s$ we introduce the normalized effective surface mean squared
displacement
\begin{equation}
\left<\mathbf{r}^2(t)\right>_s^{\mathrm{norm}}\equiv\frac{1}{N_s(t)}
\left<\mathbf{r}^2(t)\right>_s.
\end{equation}
This quantity can be interpreted as the surface mean squared displacement
covered by an individual particle that effectively stays on the surface and
does not fully escape to the bulk. At long times this quantity has the
limiting form
\begin{equation}
\left<\mathbf{r}^2(t)\right>_s^{\mathrm{norm}}\sim4D_bt.
\end{equation}
The long time diffusion corrected for the number of escaping particles
displays normal diffusion, albeit with the \emph{bulk diffusivity}.

In the following we neglect contributions from the surface diffusion
proportional to $D_s$, in order not to overburden the presentation.
The interesting behavior is due to the bulk mediation with weight
$D_b$. We quantify the motion in terms of fractional order moments,
before embarking for the surface propagator.

\subsection*{Fractional order moments}

We now derive an exact expression for the $q$-th order moments ($0<q<2$)
\begin{equation}
\label{qth}
\left<|\mathbf{r}|^q(t)\right>_s=\int|\mathbf{r}|^qn_s(\mathbf{r},t)d^2
\mathbf{r}.
\end{equation}
To this end we utilize the following integral on the plane:
\begin{widetext}
\begin{equation}
\int\Big(1-\cos(\mathbf{k}\cdot\mathbf{r})\Big)\frac{dk}{|\mathbf{k}|^{2+q}}
=2\pi\int_0^{\infty}\Big(1-J_0(kr)\Big)\frac{dk}{k^{1+q}}
=2\pi r^q\int_0^{\infty}\Big(1-J_0(z)\Big)\frac{dz}{z^{1+q}},
\end{equation}
\end{widetext}
where we used polar coordinates $r=|\mathbf{r}|$ and $k=|\mathbf{k}|$
corresponding to the two-dimensional vectors $\mathbf{r}=x\mathbf{e}_x+y
\mathbf{e}_y$ and $\mathbf{k}=k_x\mathbf{e}_x+k_y\mathbf{e}_y$. Thus we
identify
\begin{equation}
r^q=K(q)\int\Big(1-\cos(\mathbf{k}\cdot\mathbf{r})\Big)\frac{d\mathbf{k}}{
k^{2+q}},
\end{equation}
with the definition
\begin{widetext}
\begin{equation}
K(q)=\left(2\pi\int_0^{\infty}\Big(1-J_0(z)\Big)\frac{dz}{z^{1+q}}\right)^{-1}
=\frac{2^q}{\pi^2}\sin\left(\frac{\pi q}{2}\right)\left[\Gamma\left(1+\frac{q}{
2}\right)\right]^2.
\end{equation}
With this trick we can rephrase the $q$th order moment (\ref{qth}) as follows,
\begin{equation}
\langle r^q(t)\rangle_s=K(q)\int\int
\left(1-\cos\left(\mathbf{k}\cdot\mathbf{r}\right)\right)n_s(\mathbf{r},t)
\frac{d\mathbf{k}}{k^{2+q}}d\mathbf{r}=K(q)\int\left[\int n_s(\mathbf{r},t)
d\mathbf{r}-\int\cos\left(\mathbf{k}\cdot\mathbf{r}\right)n_s(\mathbf{r},t)
d\mathbf{r}\right]\frac{d\mathbf{k}}{k^{2+q}}.
\end{equation}
\end{widetext}
The integral over $d^2\mathbf{r}$ of the surface density is but the number
of surface particles, such that
\begin{equation}
\langle r^q(t)\rangle_s=K(q)\int\frac{d^2\mathbf{k}}{k^{2+q}}\Big[N_s(t)-
\mathrm{Re}\{n_s(\mathbf{k},t)\}\Big],
\end{equation}
where we replaced the Fourier cosine transform of $n_s(\mathbf{r},t)$ by
the real part of the exponential Fourier transform.

With the Fourier-Laplace transform (\ref{result_fl}) of the surface propagator
with $D_s$ set to zero, for the Laplace transform of the $q$th order moment we
obtain
\begin{widetext}
\begin{eqnarray}
\nonumber
\langle r^q(s)\rangle_s&=&K(q)\int\left[n_s(\mathbf{k}=0,s)-n_s(\mathbf{k},s)
\right]\frac{d\mathbf{k}}{k^{2+q}}=2\pi K(q)\left[\frac{1}{s+\chi\sqrt{s}}-
\frac{1}{s+\chi\sqrt{s+D_bk^2}}\right]\frac{dk}{k^{1+q}}\\
\nonumber
&=&2\pi K(q)\frac{\chi}{s+\chi\sqrt{s}}\int_0^{\infty}\frac{\sqrt{s+D_bk^2}-
\sqrt{s}}{s+\chi\sqrt{s+D_bk^2}}\frac{dk}{k^{1+q}}\\
\nonumber
&=&2\pi K(q)\frac{\chi D_b}{s+\chi\sqrt{s}}\int_0^{\infty}\frac{1}{\left(\sqrt{s}
+\sqrt{s+D_bk^2}\right)\left(s+\chi\sqrt{s+D_bk^2}\right)}\frac{dk}{k^{q-1}}\\
\nonumber
&=&2\pi K(q)\frac{\chi^2}{s+\chi\sqrt{s}}\left(\frac{D_b}{s}\right)^{q/2}\int_0
^{\infty}\frac{1}{\left(\chi+\chi\sqrt{1+y^2}\right)\left(\sqrt{s}+\chi\sqrt{
1+y^2}\right)}\frac{dy}{y^{q-1}}\\
&=&2\pi K(q)\frac{\chi D_b^{q/2}}{s^{(1+q)/2}\left(s-\chi^2\right)}\Big\{
I_1(q)-I_2(q)\Big\},
\label{qth_1}
\end{eqnarray}
\end{widetext}
where on the way we introduced the substitution $y=\sqrt{D_b/s}k$. The two
integrals $I_i$ are defined by
\begin{equation}
\label{int1}
I_1(q)=\int_0^{\infty}\frac{y^{1-q}}{1+\sqrt{1+y^2}}dy
\end{equation}
and
\begin{equation}
\label{int2}
I_2(q)=\int_0^{\infty}\frac{y^{1-q}}{\sqrt{s}/\chi+\sqrt{1+y^2}}dy.
\end{equation}
Now we analyze the temporal behavior of the $q$th order moment at short and
long times.

\subsubsection{Long time behavior}

At long times $t\gg1/\chi^2$ (or $s/\chi^2\ll 1$) we see that
\begin{equation}
I_1(q)-I_2(q)\sim\int_0^{\infty}\left(\frac{1}{1+\sqrt{1+y^2}}-\frac{1}{\sqrt{
1+y^2}}\right)\frac{dy}{y^{q-1}}<0,
\end{equation}
and
\begin{equation}
\langle r^q(s)\rangle_s\simeq\frac{1}{s^{(1+q)/2}},
\end{equation}
corresponding to the time evolution
\begin{equation}
\langle r^q(t)\rangle_s\simeq t^{(q-1)/2}.
\end{equation}
Taking normalization by the number of surface particles into account, we find
\begin{equation}
\label{long_qth}
\langle r^q(s)\rangle_s^{\mathrm{norm}}=\frac{\langle r^q(s)\rangle_s}{
N_s(t)}\simeq t^{q/2}.
\end{equation}
That is, at long times the surface diffusion exhibits normal scaling
behavior.

\subsubsection{Short time behavior}

The more interesting case is the short time behavior corresponding to the limit
$t\ll1/\chi^2$ (or $s/\chi^2\gg1$). Here we consider three separate cases:

(i) The case $1<q<2$: Since both integrals $I_1$ and $I_2$ converge, we may
simply neglect $I_2(q)$. Then
\begin{equation}
\langle r^q(s)\rangle_s\simeq s^{-(3+q)/2},
\end{equation}
such that after Laplace inversion we find
\begin{equation}
\langle r^q(t)\rangle_s\simeq t^{(q+1)/2}.
\end{equation}

(ii) The case $0<q<1$: Now we should take into account both integrals,
\begin{eqnarray}
\nonumber
I_1(q)-I_2(q)&\sim&\\
&&\hspace*{-2.4cm}
\frac{\sqrt{s}}{\chi}\int_0^{\infty}\frac{y^{1-q}dy}{\left(1+
\sqrt{1+y^2}\right)\left(\sqrt{s}/\chi+\sqrt{1+y^2}\right)}.
\end{eqnarray}
To estimate the main contribution from this difference, we split it into three
parts, namely
\begin{eqnarray}
\nonumber
I_1(q)-I_2(q)&\sim&\\
&&\hspace*{-2.4cm}
\frac{\sqrt{s}}{\chi}\left\{\int_0^1\ldots dy+\int_1^{\sqrt{s}/
\chi}\ldots dy+\int_{\sqrt{s}/\chi}^{\infty}\ldots dy\right\},
\end{eqnarray}
and evaluate each contribution separately. We find
\begin{subequations}
\begin{equation}
\int_0^1\ldots dy\simeq\frac{\chi}{\sqrt{s}},
\end{equation}
and then
\begin{equation}
\label{app_int1}
\int_1^{\sqrt{s}/\chi}\ldots dy\sim\int_1^{\sqrt{s}/\chi}\frac{1}{y\sqrt{s}/
\chi}\frac{dy}{y^{q-1}}\sim\left(\frac{\chi}{\sqrt{s}}\right)^q>\frac{\chi}{s}.
\end{equation}
Finally,
\begin{equation}
\label{app_int2}
\int_{\sqrt{s}/\chi}^{\infty}\frac{dy}{y^{1+q}}\simeq\left(\frac{\chi}{\sqrt{s}}
\right)^q.
\end{equation}
\end{subequations}
Thus the main contribution come from Eqs.~(\ref{app_int1}) and (\ref{app_int2}),
and
\begin{equation}
I_1(q)-I_2(q)\simeq s^{(1-q)/2}.
\end{equation}
With this estimate the Laplace transform of the $q$th order moment,
Eq.~(\ref{qth_1}), has the
leading order behavior
\begin{equation}
\langle r^q(s)\rangle_s\simeq\frac{s^{(1-q)/2}}{s^{(3+q)/2}}\simeq\frac{1}{
s^{1+q}}.
\end{equation}
This corresponds to
\begin{equation}
\langle r^q(t)\rangle_s\simeq t^q
\end{equation}
in the time domain.

(iii) The case $q=1$: This special case requires some care. We start with
the substitution $y=\tan\phi$ in Eqs.~(\ref{int1}) and (\ref{int2}).
Then the integrals $I_i$ become
\begin{equation}
I_1(1)=\int_0^{\pi/2}\frac{d\phi}{\cos\phi(1+\cos\phi)},
\end{equation}
and
\begin{equation}
I_2(1)=\int_0^{\pi/2}\frac{d\phi}{\cos\phi(1+\sqrt{s}\cos\phi/\chi)}.
\end{equation}
We can now rewrite Eq.~(\ref{qth_1}) in the form
\begin{eqnarray}
\nonumber
\langle r(s)\rangle_s&=&2\pi K(1)\frac{\chi D_b^{1/2}}{s(s-\chi^2)}\\
&\times&\left(\frac{\sqrt{s}}{\chi}\int_0^{\pi/2}\frac{d\phi}{1+\sqrt{s}
\cos\phi/\chi}-1\right),
\label{help}
\end{eqnarray}
where we used \cite{prudnikov}
\begin{equation}
\int_0^{\pi/2}\frac{d\phi}{1+\cos\phi}=1.
\end{equation}
For $s/\chi^2<1$ the integral in the parenthesis of Eq.~(\ref{help}) becomes
\cite{prudnikov}
\begin{equation}
\int_0^{\pi/2}\frac{d\phi}{1+\sqrt{s}\cos\phi/\chi}=\frac{2}{\sqrt{1-s/\chi^2}}
\arctan\sqrt{\frac{1-\sqrt{s}/\chi}{1+\sqrt{s}/\chi}}.
\end{equation}
Thus, for Eq.~(\ref{help}) we find
\begin{eqnarray}
\nonumber
\langle r(s)\rangle_s&=&2\pi K(1)\frac{\chi D_b^{1/2}}{s(s-\chi^2)}\\
&&\hspace*{-1.8cm}
\times\left(\frac{2\sqrt{s}}{\chi}\frac{1}{\sqrt{1-s/\chi^2}}
\arctan\sqrt{\frac{1-\sqrt{s}/\chi}{1+\sqrt{s}/\chi}}-1\right),
\end{eqnarray}
so that we find the $s\to0$ behavior
\begin{equation}
\langle r(s)\rangle_s\sim\frac{D_b^{1/2}}{\chi s}.
\end{equation}
After Laplace inversion,
\begin{equation}
\langle r(t)\rangle_s\sim\frac{D_b^{1/2}}{\chi},\,\,\,\mbox{and}\,\,\,
\langle r(t)\rangle_s^{\mathrm{norm}}\sim(D_bt)^{1/2}
\end{equation}
at long times, $t\gg1/\chi^2$, consistent with Eq.~(\ref{long_qth}).
Conversely, for $s/\chi^2>1$ we employ \cite{prudnikov}
\begin{eqnarray}
\nonumber
\int_0^{\pi/2}\frac{d\phi}{1+\sqrt{s}\cos\phi/\chi}&=&\frac{1}{\sqrt{s/\chi^2-
1}}\\
&&\hspace*{-1.8cm}\times
\ln\frac{\sqrt{s/\chi^2-1}+\sqrt{s}/\chi-1}{\sqrt{s/\chi^2-1}+1-\sqrt{s}/\chi}.
\end{eqnarray}
Thus we find
\begin{equation}
\langle r(s)\rangle_s=2\pi K(1)\frac{\chi D_b^{1/2}}{s(s-\chi^2)}\left[\frac{
\sqrt{s}/\chi}{\sqrt{s/\chi^2-1}}\ln\left(\frac{2\sqrt{s}}{\chi}\right)-1
\right].
\end{equation}
We thus obtain the limiting form at large $s$
\begin{equation}
\langle r(s)\rangle_s\sim\frac{\chi D_b^{1/2}}{s^2}\ln\left(\frac{2\sqrt{s}}{
\chi}\right).
\end{equation}
Back-transformed this results in
\begin{equation}
\langle r(t)\rangle_s\approx\langle r^q(s)\rangle_s^{\mathrm{norm}}\sim t\ln t
\end{equation}
at short times $t\ll1/\chi^2$.

Summarizing our results for $q$th order moments, at short times $t\ll1/\chi^2$
the effective surface diffusion exhibits anomalous scaling: the $q$th order
moment of the radius scales like $t^q$ for $0<q<1$ and $t^{(q+1)/2}$ for $1<q<2$,
while the first moment includes a logarithmic contribution, $t\ln t$, consistent
with the earlier results in Ref.~\cite{bychuk1}. At long times $t\gg1/\chi^2$
the $q$th order moments scale normally with time, proportional to $t^{q/2}$.

\section{Surface propagator}

We now turn to the behavior of the surface propagator.
If we neglect surface diffusion (i.e., $D_s=0$) the surface propagator
from Eq.~(\ref{result_fl}) becomes
\begin{equation}
n_s(\mathbf{k},s)=\frac{1}{s+\chi\sqrt{s+D_bk^2}}.
\end{equation}
We perform an inverse Laplace transformation along the Bromwich path:
\begin{widetext}
\begin{eqnarray}
\nonumber
n_s(\mathbf{k},t)&=&\int_{\mathrm{Br}}\frac{\exp(st)}{s+\chi\sqrt{s+D_bk^2}}\frac{
ds}{2\pi i}=\int_{\mathrm{Br}}\frac{\exp(s\chi^2t)}{s+\sqrt{s+\kappa^2}}\frac{
ds}{2\pi i}=e^{-\kappa^2\chi^2t}\int_{\mathrm{Br}}\frac{\exp(s\chi^2t)}{s+
\sqrt{s}-\kappa^2}\frac{ds}{2\pi i}\\
\nonumber
&&\hspace*{-1.8cm}
=e^{-\kappa^2\chi^2t}\int_{\mathrm{Br}}\frac{\exp(s\chi^2t)}{\left(\sqrt{s}
+\frac{1}{2}+\sqrt{\kappa^2+\frac{1}{4}}\right)\left(\sqrt{s}+\frac{1}{2}-\sqrt{
\kappa^2+\frac{1}{4}}\right)}\frac{ds}{2\pi i}\\
\nonumber
&&\hspace*{-1.8cm}
=\frac{\exp(-\kappa^2\chi^2t)}{2\sqrt{\kappa^2+1/4}}\left(\int_{\mathrm{Br}}
\frac{\exp(s\chi^2t)}{\sqrt{s}+\frac{1}{2}-\sqrt{\kappa^2+\frac{1}{4}}}\frac{ds
}{2\pi i}-\int_{\mathrm{Br}}\frac{\exp(s\chi^2t)}{\sqrt{s}+\frac{1}{2}+\sqrt{
\kappa^2+\frac{1}{4}}}\frac{ds}{2\pi i}\right)\\
&&\hspace*{-1.8cm}
=\frac{\exp(-\kappa^2\chi^2t)}{2\sqrt{\kappa^2+1/4}}\left(\mathscr{L}^{-1}
\left\{\frac{1}{\sqrt{s}-b};s\to t'\right\}_{b=\sqrt{\kappa^2+1/4}-1/2,t'=t
\chi^2}-\mathscr{L}^{-1}\left\{\frac{1}{\sqrt{s}+a};s\to t'\right\}_{a=1/2+
\sqrt{\kappa^2+1/4},t'=t\chi^2}\right)
\end{eqnarray}
where on the way we introduced the substitutions $s'=s/\chi^2$ and $\kappa^2=
D_bk^2/\chi^2$. With
\begin{equation}
\mathscr{L}^{-1}\left\{\frac{1}{\sqrt{s}+a};s\to t\right\}=\frac{1}{\sqrt{\pi
t}}-ae^{a^2t}\mathrm{erfc}\left(a\sqrt{t}\right)
\end{equation}
we finally arrive at the Fourier transform of the surface propagator,
\begin{eqnarray}
\nonumber
n_s(\mathbf{k},t)&=&\frac{1}{2\sqrt{\kappa^2+1/4}}\left\{\left[\frac{1}{2}+\sqrt{
\kappa^2+\frac{1}{4}}\right]\exp\left(\left[\frac{1}{2}+\sqrt{\kappa^2+\frac{1}{
4}}\right]t\chi^2\right)\mathrm{erfc}\left(\left[\frac{1}{2}+\sqrt{\kappa^2+
\frac{1}{4}}\right]\chi\sqrt{t}\right)\right.\\
&&+\left.
\left(\sqrt{\kappa^2+\frac{1}{4}}-\frac{1}{2}\right)\exp\left(\left[\frac{1}{
2}-\sqrt{\kappa^2+\frac{1}{4}}\right]t\chi^2\right)\mathrm{erfc}\left(\left[
\frac{1}{2}-\sqrt{\kappa^2+\frac{1}{4}}\right]\chi\sqrt{t}\right)\right\}.
\label{kt_prop}
\end{eqnarray}
\end{widetext}
In compact form,
\begin{eqnarray}
\nonumber
n_s(\mathbf{k},t)&=&\frac{1}{\alpha+\beta}\left\{\alpha e^{\alpha\chi^2t}
\mathrm{erfc}\left(\alpha\chi\sqrt{t}\right)\right.\\
&&\left.+\beta e^{-\beta\chi^2t}\mathrm{erfc}\left(-\beta\chi\sqrt{t}\right)
\right\},
\label{compact}
\end{eqnarray}
with
\begin{equation}
\alpha=\sqrt{\kappa^2+\frac{1}{4}}+\frac{1}{2},\quad
\beta=\sqrt{\kappa^2+\frac{1}{4}}-\frac{1}{2}.
\end{equation}

We now calculate the surface propagator $n_s(\mathbf{r},t)$ in the limits of
short and long times.

\subsection{Short time behavior}

At short times $t\ll\chi^{-2}$ we distinguish between the central part of the
propagator and its wings. Starting with the central part, we thus focus on
distances $r\ll(D_bt)^{1/2}$. Due to $\kappa^2=D_bk^2/\chi^2$, this means
that we consider $\kappa\chi\sqrt{t}\gg1$, and therefore $\kappa^2\gg1$.
Consequently we find $\alpha\approx\beta\approx\kappa$, $\alpha\chi\sqrt{t}
\approx\beta\chi\sqrt{t}\approx\kappa\chi\sqrt{t}\gg1$, and thus
\begin{equation}
\mathrm{erfc}\left(\alpha\chi\sqrt{t}\right)\sim\frac{\exp(-\alpha^2\chi^2t)}{
\sqrt{\pi}\alpha\chi\sqrt{t}}.
\end{equation}
Then, the first term in the curly brackets of Eq.~(\ref{compact}) is
(note that $\alpha^2-\alpha=\kappa^2$)
\begin{equation}
\alpha e^{\alpha\chi^2 t}\mathrm{erfc}\left(\alpha\chi\sqrt{t}\right)\approx
\frac{1}{\sqrt{\pi\chi^2 t}}e^{-\kappa^2\chi^2 t}.
\end{equation}
The error function in the second term of Eq.~(\ref{compact}) is approximately 2.
Thus, the second term in the curly brackets of Eq.~(\ref{compact}) prevails,
and we obtain
\begin{equation}
n_s(\mathbf{k},t)\sim\exp\left(-\kappa\chi^2t\right).
\end{equation}
The Fourier inversion is accomplished with the help of the integral
\cite{prudnikov}
\begin{equation}
\int_0^{\infty}xe^{-px}J_0(cx)dx=\frac{p}{(p^2+c^2)^{3/2}},
\end{equation}
and we obtain
\begin{eqnarray}
\nonumber
n_s(\mathbf{r},t)&\sim&\int e^{-i\mathbf{k}\cdot\mathbf{r}}n_s(\mathbf{k},t)
\frac{d\mathbf{k}}{4\pi^2}\\
&\sim&\int_0^{\infty}e^{-\chi\sqrt{D_b}kt}J_0(kr)\frac{kdk}{2\pi}
\end{eqnarray}
such that
\begin{equation}
\label{cauchy}
n_s(\mathbf{r},t)\sim\frac{\chi D_b^{1/2}t}{2\pi\left(r^2+\chi^2D_bt^2
\right)^{3/2}}.
\end{equation}
This is exactly the two-dimensional Cauchy distribution obtained by Bychuk
and O'Shaugnessy from scaling arguments \cite{bychuk}. We see that indeed
radius and time are coupled linearly in superdiffusive fashion, $r\sim t$.
However, as we will see this Cauchy form holds only for the central part
of the surface propagator. It assumes steeper wings such that at all times
any moment of the propagator exists. We are quantifying these wings below.

The number of particles populating this cental Cauchy domain of the propagator,
i.e., within a circle of radius $(D_bt)^{1/2}$ is
\begin{equation}
\int^{(D_bt)^{1/2}}n_s(\mathbf{r},t)d\mathbf{r}\sim1-\chi\sqrt{t},
\end{equation}
i.e., at sufficiently short times a major portion of the particles are
contained in the Cauchy domain. Indeed, these particles show the
superdiffusive mean squared displacement
\begin{equation}
\langle\mathbf{r}^2\rangle_s=\int^{(D_bt)^{1/2}}\mathbf{r}^2n_s(\mathbf{r},t)
d\mathbf{r}\approx\chi D_bt^{3/2},
\end{equation}
consistent with result (\ref{msd_asymp}) for $D_s=0$, the prefactor differing
by approximately a factor $3/2$.

The outer part of the propagator at short times is found for $r\gg(D_bt)^{1/2}$.
This region can be divided into two subregions, namely, $\kappa^2\gg1$
corresponding to $D_bt\ll r^2\ll D_b/\chi^2$, and $\kappa^2\ll1$ corresponding
to $r^2\gg D_b/\chi^2$. However, it is easy to check that in both of these
sub-regions we have $\alpha\chi\sqrt{t}\ll1$ and $\beta\chi\sqrt{t}\ll1$,
due to the combination of the conditions of short times ($\chi\sqrt{t}\ll1$)
and large radii ($\kappa\chi\sqrt{t}\ll1$). Thus both error functions in
Eq.~(\ref{compact}) are expanded at small values of their arguments,
\begin{equation}
\mathrm{erfc}(\pm z)\sim1\mp\frac{2}{\sqrt{\pi}}z\pm\frac{2}{3\sqrt{\pi}}z^3
+\ldots,
\end{equation}
with $z=\alpha\chi\sqrt{t}<1$ and $z=\beta\chi\sqrt{t}<1$, respectively. Our
result is
\begin{equation}
n_s(\mathbf{k},t)\sim1-\frac{2}{\sqrt{\pi}}\chi\sqrt{t}-\frac{2}{3\sqrt{\pi}}
k^2D_b\chi t^{3/2}
\end{equation}
plus terms of order $k^4$ and higher. We conclude that at the wings the central
Cauchy distribution is truncated by Gaussian tails whose dispersion grows like
$t^{3/2}$.

\subsection{Long time behavior}

This is the limit $t\gg1/\chi^2$. As this implies $\alpha\chi\sqrt{t}\gg1$ we
may expand the error function as
\begin{equation}
\mathrm{erfc}(\alpha\chi\sqrt{t})\sim\frac{\exp(-\alpha^2\chi^2t)}{\sqrt{\pi}
\alpha\chi\sqrt{t}}.
\end{equation}
We consider the shape of the propagator in the main part excluding the 
region near the origin, i.e, we have
\begin{equation}
r^2\gg D_b\sqrt{t}/\chi\gg D_b/\chi^2.
\end{equation}
In this domain we may use $\kappa\ll1$, $\alpha\approx1$, and $\kappa^2\chi
\sqrt{t}\ll1$. Since $\beta=\sqrt{\kappa^2+1/4}-1/2\approx\kappa^2$, we have
$\beta\chi\sqrt{t}\ll1$, and thus the second error function becomes
\begin{equation}
\mathrm{erfc}(-\beta\chi\sqrt{t})\approx1.
\end{equation}
The surface propagator thus assumes the shape (recall that $\alpha^2-\alpha
=\kappa^2$)
\begin{eqnarray}
\nonumber
n_s(\mathbf{k},t)&\sim&\frac{1}{\alpha}\left(\alpha e^{\alpha\chi^2t}\frac{
\exp(-\alpha^2\chi^2t)}{\sqrt{\pi}\alpha\chi\sqrt{t}}+\kappa^2\exp\left(-\beta
\chi^2t\right)\right)\\
\nonumber
&\sim&\frac{\exp(-\kappa^2\chi^2t)}{\sqrt{\pi}\chi\sqrt{t}}+\kappa^2e^{-\kappa^2
\chi^2t}\\
&\sim&\frac{\exp(-\kappa^2\chi^2t)}{\sqrt{\pi}\chi\sqrt{t}}.
\end{eqnarray}
Equivalently, this means that
\begin{equation}
n_s(\mathbf{k},t)\sim\frac{\exp(-k^2D_bt)}{\sqrt{\pi}\chi\sqrt{t}}.
\end{equation}
Inverse Fourier transform produces the propagator at long times,
\begin{eqnarray}
\nonumber
n_s(\mathbf{r},t)&\sim&\int e^{-i\mathbf{k}\cdot\mathbf{r}}n_s(\mathbf{k},t)
\frac{d\mathbf{k}}{4\pi^2}\\
&\sim&\frac{1}{\chi\sqrt{\pi t}}\int_0^{\infty}J_0(kr)e^{-k^2D_bt}\frac{kdk}{
2\pi}.
\end{eqnarray}
From this we obtain the Gaussian form
\begin{equation}
\label{gauss}
n_s(\mathbf{r},t)\sim\frac{1}{\chi\sqrt{\pi t}}\frac{\exp(-r^2/[4D_bt])}{4
\pi D_bt}.
\end{equation}
The number of particles contained in this part of the long-time propagator
follows from
\begin{equation}
\int_{r^2\ge D_b\sqrt{t}/\chi}n_s(\mathbf{r},t)d\mathbf{r}\sim\frac{1}{\chi
\sqrt{\pi t}}.
\end{equation}
Comparison to Eq.~(\ref{numb_long}) shows that these are virtually all
particles still on the surface. Thus the Gaussian (\ref{gauss}) indeed
dominates the behavior at long times. This statement is also corroborated
from calculation of the surface mean squared displacement from Eq.~(\ref{gauss})
in the domain $r^2\ge D_b\sqrt{t}/\chi$:
\begin{eqnarray}
\nonumber
\langle\mathbf{r}^2\rangle_s&\sim&\int_{r^2\ge D_b\sqrt{t}/\chi}r^2n_s(\mathbf{
r},t)d\mathbf{r}\\
&\sim&\frac{4D_bt}{\chi\sqrt{\pi t}}\left[1+o\left(\frac{1}{t\chi^2}\right)
\right].
\end{eqnarray}
Again, up to a small correction, this term coincides with our exact result for
the long time behavior, Eq.~(\ref{momlong}).

\section{Conclusions}

Complementing our earlier studies of bulk mediated surface diffusion on a
cylinder \cite{bmsdlong}, we presented a strictly analytical treatment of
the bulk mediated surface diffusion problem on a flat surface. While our
analysis corroborates previous scaling results of superdiffusion and the
Cauchy form of the surface propagator, we find corrections to this behavior
sufficiently far away from the origin, and at long times. Thus, the Cauchy
domain is rectified by a Gaussian decay of the extremities of the propagator
which we derive explicitly. As a consequence finite spatial moments of any
order remain finite. At long times a Gaussian regime takes over, assembling
virtually all particles.

Our analysis is based on the exact representation of the surface propagator
in Fourier-Laplace space from which systematic expansions are obtained. For
the number of surface particles we find an initial stagnation, followed at
longer times by a decrease proportional to $1/\sqrt{t}$. The mean squared
displacement $\langle\mathbf{r}^2\rangle_s$ turns from the initial
superdiffusive behavior proportional to $t^{3/2}$ to a subdiffusive form 
$\langle\mathbf{r}^2\rangle_s\simeq t^{1/2}$. The latter corresponds to
normal diffusion, albeit with the bulk diffusivity, if normalized to the
decay of the surface particles. Our analysis is complemented with the
derivation of fractional order moments for the surface propagation.

\begin{acknowledgments}
RM thanks the Academy of Finland for financial support within the FiDiPro
scheme.
\end{acknowledgments}

\appendix

\section{Alternative derivation of the surface density}

In this appendix we derive the L{\'e}vy flight behavior recovered in the
main part of the text based on a complementary approach developed in
Refs.~\cite{lomholt07,irwin09,lomholt09}. We start with the coupled diffusion
equation along the surface (equivalent to Eq.~(\ref{surf_diff})) as
\begin{widetext}
\begin{eqnarray}
\frac{\partial n_s}{\partial t}&=&D_s\left(\frac{\partial^2}{\partial x^2}
+\frac{\partial^2}{\partial y^2}\right)n_s-\frac{1}{\tau_{\rm des}}n_s+
\frac{1}{\tau_{\rm des}}\int d x'\int d y'\int_0^t d t'\, W_{\rm bulk}(x-x',
y-y',t-t')n_s(x',y',t'),
\label{eq:Astart}
\end{eqnarray}
\end{widetext}
where the last term represents rebinding events of particles that previously
desorbed from the surface. The kernel $W_{\rm bulk}(x,y,t)$ in the
convolution is the probability density to attach at $(x,y)$ at time $t$
after desorption from the origin at time zero. Together, the last two terms
of Eq.~(\ref{eq:Astart}) represent jumps along the surface mediated by the
bulk. Note that we have implicitly assumed that all particles start at the
surface at $t=0$. Otherwise an additional term would be needed to represent
particles binding time to the surface for the first time.

To obtain $W_{\rm bulk}$ we need to solve the bulk diffusion equation,
Eq.~(\ref{bulk_diff}), with a single particle released from the boundary at
the origin at time zero. This can be achieved by choosing a boundary condition
\begin{equation}
D_b\left.\frac{\partial n_b}{\partial z}\right|_{z=0}=\frac{1}{\mu\tau_{\rm
des}}n_b|_{z=0}-\delta(x)\delta(y)\delta(t).
\label{eq:Aboundary}
\end{equation}
This equation corresponds to the current of particles into the surface,
the second term on the right hand side representing the initial
release. Supplementing this by boundary conditions of vanishing density at
infinity, $n_b\to 0$ as $|x|,|y|,z\to \infty$, and the initial condition
of finding no particles in the bulk, $n_b(x,y,z,t=0)=0$, we obtain $W_{\rm
bulk}$ from the solution in terms of the flux into the boundary after the
initial release: $W_{\rm bulk}(x,y,t)=n_b(x,y,z=0,t)/(\mu\tau_{\rm des})$.

To solve Eq.~(\ref{bulk_diff}) we again Fourier transform along $x$ and $y$
($x\to k_x$ and $y\to k_y$) and Laplace transform in time ($t\to s$). Together
with the initial condition and the boundary conditions at infinity it is
easily found that
\begin{equation}
n_b(k_x,k_y,z,s)=e^{-z\sqrt{k^2+s/D_b}}n_b(k_x,k_y,z=0,s),\label{eq:nbsol}
\end{equation}
where $k^2=k_x^2+k_y^2$. Insertion of Eq.~(\ref{eq:nbsol}) into
Eq.~(\ref{eq:Aboundary}) leads to a bulk density at the boundary which is
\begin{equation}
n_b(k_x,k_y,z=0,s)=\frac{1}{(\mu\tau_{\rm des})^{-1}+D_b\sqrt{k^2+s/D_b}},
\end{equation}
Multiplication by the rate constant $1/(\mu\tau_{\rm des})$ then leads to
\begin{equation}
W_{\rm bulk}(k_x,k_y,s)=\frac{1}{1+\mu\tau_{\rm des}D_b\sqrt{k^2+s/D_b}}.
\label{eq:Wbulk}
\end{equation}

We are now ready to obtain the surface density $n_s$ in Fourier-Laplace space.
From Eq.~(\ref{eq:Astart}) with the initial condition $n_s(x,y,t=0)=n_0\delta
(x)\delta(y)$ we obtain
\begin{equation}
n_s(k_x,k_y,s)=\frac{n_0}{s+D_s k^2+\tau_{\rm des}^{-1}[1-W_{\rm bulk}(k_x,k_y,
s)]},
\label{eq:A6}
\end{equation}
in which Eq.~(\ref{eq:Wbulk}) can be directly inserted. To obtain the
expression in Eq.~(\ref{result_fl}) of the main text one takes the limit
of the desorption rate constant $\tau_{\rm des}^{-1}$ and adsorption rate
constant $1/(\mu\tau_{\rm des})$ going to infinity while keeping their ratio
$\mu$ fixed, which leads to
\begin{equation}
\tau_{\rm des}^{-1}[1-W_{\rm bulk}(k_x,k_y,s)]\to \mu D_b\sqrt{k^2+s/D_b}.
\end{equation}
Together with $\chi=\mu\sqrt{D_b}$ and $n_0=1$ this makes Eq.~(\ref{eq:A6})
equivalent to Eq.~(\ref{result_fl}).

As a byproduct of the approach of this appendix we remark that we can
now obtain the distribution of bulk mediated jump lengths. These are
given by $\lambda(r=\sqrt{x^2+y^2})=2\pi r W_{\rm bulk}(x,y,s=0)$. From
Eq.~(\ref{eq:Wbulk}) we find [performing the angular part of the inverse
Fourier transform similarly to Eq.~(\ref{cauchy})]
\begin{widetext}
\begin{equation}
\lambda(r)=2\pi r\int_0^\infty\frac{k d k}{2\pi}\frac{J_0(k r)}{1+
\mu\tau_{\rm des}D_b k}
=\frac{1}{\mu \tau_{\rm des} D_b}-\frac{\pi r}{2(\mu \tau_{\rm des}
D_b)^2}\bigg[{\bf H}_0\left(\frac{r}{\mu \tau_{\rm des} D_b}\right)
-Y_0\left(\frac{r}{\mu \tau_{\rm des} D_b}\right)\bigg],
\end{equation}
where ${\bf H}_\nu$ are the Struve functions and $Y_\nu$ are the Bessel
functions of the second kind. If we expand $\lambda(r)$ asymptotically
at large $r$ we obtain a power law
\begin{equation}
\lambda(r)\sim \frac{\mu \tau_{\rm des} D_b}{r^2}.
\end{equation}
Thus we see that the Cauchy propagator (\ref{cauchy}), arising for $D_s=0$
in the short time regime where almost all particles are still bound on the
surface, is connected with a L{\'e}vy flight behavior based on a power law jump
length distribution.

\section{Derivation of the coupling equation (\ref{bound})}
\label{app1}

We start by rewriting Eq.~(\ref{bulk_fl}) in the form
\begin{equation}
\label{app1_1}
n_b(x,y,z,t)=\frac{\mu z}{(4\pi D_b)^{3/2}}\int_0^td\tau\int dx'\int dy'\frac{
n_s(x',y',t-\tau)}{\tau^{5/2}}\exp\left(-\frac{(x-x')^2+(y-y')^2+z^2}{4D_b\tau}
\right).
\end{equation}
We know that the Gaussian is a limiting representation of the $\delta$ function:
\begin{equation}
\lim_{\tau\to0}\frac{1}{\sqrt{4\pi D_b\tau}}\exp\left(-\frac{(x-x')^2}{4D_b\tau}
\right)=\delta(x-x').
\end{equation}
Let us now introduce the new variable $\tau'=\tau/z^2$, such that relation
(\ref{app1_1}) turns into
\begin{equation}
n_b(x,y,z,t)=\frac{\mu z}{(4\pi D_b)^{3/2}}\int_0^{t/z^2}z^2d\tau'\int dx'\int
dy'\frac{n_s(x',y',t-\tau'z^2)}{z^5(\tau')^{5/2}}\exp\left(-\frac{(x-x')^2}{4D_b
\tau'z^2}-\frac{(y-y')^2}{4D_b\tau'z^2}-\frac{z^2}{4D_b\tau'z^2}\right).
\end{equation}
Rearranging terms,
\begin{equation}
n_b(x,y,z,t)=\int_0^{t/z^2}d\tau'\int dx'\int dy'\frac{\mu n_s(x',y',t-\tau'z^2)
}{(\tau')^{3/2}}\frac{\exp\left(-\dfrac{(x-x')^2}{4D_b\tau'z^2}\right)}{\sqrt{4
\pi D_b\tau'z^2}}\frac{\exp\left(-\dfrac{(y-y')^2}{4D_b\tau'z^2}\right)}{\sqrt{4
\pi D_b\tau'z^2}}\frac{\exp\left(-\dfrac{1}{4D_b\tau'}\right)}{\sqrt{4 \pi D_b}}.
\end{equation}
We are now in the position to take the limit $z\to0$, producing (we omit the
prime in $\tau'$)
\begin{equation}
\lim_{z\to0}n_b(x,y,z,t)=\int_0^{\infty}d\tau\int dx'\int dy'\frac{\mu n_s(x',
y',t)}{\tau^{3/2}}\delta(x-x')\delta(y-y')\frac{\exp\left(-\dfrac{1}{4D_b\tau}
\right)}{\sqrt{4\pi D_b}},
\end{equation}
which directly reduces to the expression
\begin{equation}
\lim_{z\to0}n_b(x,y,z,t)=\mu n_s(x,y,t)\frac{1}{\sqrt{4\pi D_b}}\int_0^{\infty}
d\tau\frac{\exp\left(-\dfrac{1}{4D_b\tau}\right)}{\tau^{3/2}}
\end{equation}
Finally, with the substitution $\zeta=\tau^{-1/2}$ with the Jacobian $d\zeta=-
1/\left(2\tau^{3/2}\right)$, the remaining integral represents the normalization
of a Gaussian,
\begin{equation}
\lim_{z\to0}n_b(x,y,z,t)=\mu n_s(x,y,t)\frac{2}{\sqrt{4\pi D_b}}\int_0^{\infty}
d\zeta\exp\left(-\frac{\zeta^2}{4D_b}\right)=\mu n_s(x,y,t),
\end{equation}
and the equivalence with the coupling equation (\ref{bound}) is established.
\end{widetext}

\end{document}